\documentclass[11pt]{iopart}
\usepackage{iopams}
\usepackage{graphicx}
\usepackage[breaklinks, colorlinks=true, pdfstartview=FitV]{hyperref}

\newcommand{\hata}{{\hat a}}
\newcommand{\hatad}{{\hat a}^\dag}
\newcommand{\hatb}{{\hat b}}
\newcommand{\hatbd}{{\hat b}^\dag}

\newcommand{\hatrho}{{\hat \rho}}
\newcommand{\hatphi}{{\hat \phi}}

\newcommand{\delt}{\delta t}

\newcommand{\calA}{\mathcal{A}}
\newcommand{\calD}{\mathcal{D}}
\newcommand{\bp}{\boldsymbol{p}}
\newcommand{\bx}{\boldsymbol{x}}
\newcommand{\bA}{\boldsymbol{A}}

\begin{document}

\title[Spectral representation of the particle production]
{Spectral representation of the particle production out of equilibrium
 -- Schwinger mechanism in pulsed electric fields}

\author{Kenji Fukushima}
\address{Department of Physics, The University of Tokyo,
         7-3-1 Hongo, Bunkyo-ku, Tokyo 113-0033, Japan}
\ead{fuku@nt.phys.s.u-tokyo.ac.jp}

\begin{abstract}
  We develop a formalism to describe the particle production out of
  equilibrium in terms of dynamical spectral functions, i.e.\ Wigner
  transformed Pauli-Jordan's and Hadamard's functions.  We take an
  explicit example of a spatially homogeneous scalar theory under
  pulsed electric fields and investigate the time evolution of the
  spectral functions.  In the out-state we find an oscillatory peak in
  Hadamard's function as a result of the mixing between positive- and
  negative-energy waves.  The strength of this peak is of the linear
  order of the Bogoliubov mixing coefficient, whereas the peak
  corresponding to the Schwinger mechanism is of the quadratic order.
  Between the in- and the out-states we observe a continuous flow of
  the spectral peaks together with two transient oscillatory peaks.
  We also discuss the medium effect at finite temperature and density.
  We emphasise that the entire structure of the spectral functions
  conveys rich information on real-time dynamics including the
  particle production.
\end{abstract}

\section{Introduction}

Quantum field theory has been quite successful in describing
non-trivial contents of the vacuum and the $S$-matrix elements from
the vacuum-to-vacuum amplitude using the Lehmann-Symanzik-Zimmermann
(LSZ) reduction formula.  For static quantities the numerical
Monte-Carlo simulation of the lattice discretised theory is so
powerful that one could perform the first-principle calculation in a
non-perturbative way.  In contrast to tremendous achievements for
static quantities, the numerical machinery for solving the real-time
dynamics (or the initial value problem) has not been established
beyond the scope of the linear response theory.  The point is that one
should compute not the vacuum-to-vacuum amplitude but an expectation
value at time $t$, which involves time-evolution operators from
$-\infty$ to $t$ and also its Hermite conjugate.  It is known as the
closed-time path (CTP) formalism
\cite{Schwinger:1960qe,Keldysh:1964ud} how to deal with two
time-evolution operators with generalised Green's functions.  The
microscopic derivation of the Boltzmann equation was pioneered by
Kadanoff and Baym \cite{kadanoff1962quantum} (see also
\cite{Ivanov:2000,Kita:2006} for recent studies), in which the Wigner
transform of correlation functions translates to the distribution
function and the spectral function.

Generally speaking, the spectral functions provide us with detailed
information on physical contents in the system.  Even in the case of
equilibrated matter the spectral function represents in-medium
dispersion relations of collective excitations such as the plasmon,
the zero sound, etc (see \cite{bellac1996thermal,abrikosov2012methods}
for classical textbooks).  One can even infer the real-time properties
near equilibrium by the analytical continuation of Green's functions
once a spectral function is available.  In this way, using the
spectral function (or the imaginary part of the retarded self-energy),
one can evaluate the thermal emission rate of a pair of particle and
anti-particle (or hole in condensed matter systems)
\cite{McLerran:1984ay,Weldon:1990iw}.  Such thermal processes are
allowed in a medium where thermally excited particles are brought in.
In this kind of calculation in equilibrated matter, the translational
invariance in time needs not be violated and the ordinary field-theory
techniques are useful.

A more non-trivial example of the particle production is the process
induced by the presence of time-dependent external field.  The
pioneering work by Heisenberg and Euler \cite{Heisenberg:1935qt} has
revealed that the one-loop effective action on top of electromagnetic
background fields has an imaginary part.  This indicates that the
vacuum becomes unstable;  in other words, the particle production can
occur from the vacuum.  The vacuum permittivity has also been
formulated in the field-theoretical manner by Schwinger
\cite{Schwinger:1951nm}.  Named after his seminal work, the pair
production of particle and anti-particle from the vacuum under
electric field is commonly referred to as the Schwinger mechanism (see
\cite{Dunne:2004nc} for a comprehensive review).  This could be
regarded as a special example of the Landau-Zener effect (see
\cite{Aoki:2013} for example).

The essence for the particle production from the vacuum is concisely
represented by the Bogoliubov transformation of the
creation/annihilation operators, with which the positive- and the
negative-energy states are mixed together.  We note that the
celebrated Hawking radiation, that is the particle production under
gravitational effects, belongs to the same class of physics.  In
short, the vacuum defined in the ``in-state'' is filled with particles
and anti-particles if seen in the ``out-state'' where the observer
stands, and the Bogoliubov transformation connects the in- and the
out-states by a unitary rotation.

The Schwinger mechanism and the Hawking radiation are quantum
(tunnelling) phenomena and have been intensively studied in the
semi-classical method like the Wentzel-Kramers-Brillouin (WKB)
approximation (see \cite{Soffel:1982pm} for a review, and also
\cite{Parikh:1999mf} for the WKB formulation of the Hawking
radiation).  Although the mixing via the Bogoliubov transformation is
straightforward and the semi-classical methods appeal to our
intuition, it would be more desirable to develop a systematic
formulation in terms of the field correlators.  We believe that this
reformulation is indispensable for future progresses;  someday one
might be able to execute real-time numerical simulations, and then,
the canonical quantisation with creation/annihilation operators is not
quite compatible with numerical algorithms.  Ideally, if we can
express the Schwinger mechanism using some spectral functions in
analogy with the thermal emission rate, we could attain a unified
view of the particle production near and out of equilibrium.

Some time ago the present author has formulated the Schwinger
mechanism in a form similar to the LSZ reduction formula in
\cite{Fukushima:2009er}, which is based on preceding works
\cite{Gelis:2006yv,Gelis:2006cr}.  A variant of this formula is also
used in a recent attempt to utilise the classical statistical
approximation to simulate the Schwinger mechanism numerically
\cite{Gelis:2013oca}.  As we discuss later, though the LSZ-type
formula looks reasonable, the treatment of the in-state has some
subtlety.  If we consider the inclusive spectrum only, in fact, we can
easily derive another formula given in terms of the spectral
functions without any ambiguity.  In this case the translational
invariance in time is lost and we should handle the \textit{dynamical
  (time-dependent) spectral functions}.  It should be thus a natural
idea to look into the temporal change of the spectral functions in
accord with the quasi-particle contents affected by the time-dependent
background fields.

We stress that reformulating the problem of the particle production
makes an important building block in a timely subject;  real-time
dynamics is the key issue in various fields of physics.  In the
research of the quark-gluon plasma formation for instance, the
thermalization process is under active dispute (see
\cite{Berges:2012ks} and references therein).  Large laser facilities
are aiming to detect the production of a pair of electron and positron
and it has been discovered that the dynamically assisted Schwinger
mechanism significantly reduces the critical strength of the electric
field \cite{Schutzhold:2008pz,Dunne:2009gi}.  For precise theoretical
predictions it is strongly demanded to invent a new scheme for the
full quantum real-time simulation.  Probably, to achieve this goal,
the stochastic quantisation is one of the most promising approaches
\cite{Damgaard:1987rr}.  However, the conventional description of the
Schwinger mechanism or the Hawking radiation does not fit in with the
functional language with which the stochastic quantisation is written.
This highly motivates us to think of the spectral representation of
the particle production out of equilibrium, as is the main topic of
this work.

In this paper we will first give a detailed account of the derivation
of our formula with the spectral functions.  Then, we will investigate
the general properties of the spectral functions associated with the
in- and the out-states involving the Bogoliubov transformation.  We
can understand that the Schwinger mechanism accesses only a small
portion of the whole spectral functions.  This means, hence, that the
spectral functions contain much more information than the Schwinger
mechanism and new possibilities for a better detection might be still
buried in them.  The dynamical spectral functions thus deserve serious
investigations and we will construct them concretely for a special
case of homogeneous pulsed electric fields to dig non-trivial features
out.

\section{Particle number out of equilibrium}

Let us consider a general setup to formulate the particle production
out of equilibrium in quantum field theory.  In this paper we focus
only on a single-component complex scalar field (i.e.\ scalar QED
\cite{Kim:2008yt}) to simplify the expressions, but the generalisation
to other fields such as fermions and multi-component fields is not
difficult \cite{Dunne:1998ni}.

We require the existence of well-defined asymptotic states, namely,
the in-state at $t=-\infty$ and the out-state at $t=\infty$, where the
interactions should be turned off.  In our convention we put ``in''
and ``out'' in the subscript to refer to quantities that belong to the
in-state and the out-state, respectively.  With increasing time, thus,
the energy dispersion relation should evolve from $p_0=E_{\rm in}(\bp)$
to $p_0=E_{\rm out}(\bp)$ driven by interactions with external fields,
and we would like to compute the particle number associated with this
change.  For this purpose the expression for the inclusive spectrum is
our starting point, which is given by the number operator as
\begin{equation}
 \frac{\rmd N}{\rmd^3\bp} = \frac{1}{(2\pi)^3} \langle\hatrho_{\rm in}\,
  \hat{n}_{\rm out}(\bp) \rangle := \frac{1}{(2\pi)^3} \langle
  \hatrho_{\rm in}\, \hatad_{\rm out}(\bp)\,\hata_{\rm out}(\bp)\,
  \rangle \;.
\label{eq:spectrum}
\end{equation}
Here $\hatrho_{\rm in}$ represents the density matrix that
characterises the in-state.  If we choose it to be a pure state of
\textit{the initial vacuum},
$\hatrho_{\rm in}=|0_{\rm in}\rangle\langle 0_{\rm in}|$, there is no
contribution to \eref{eq:spectrum} from the initial state.  Then
\eref{eq:spectrum} counts the number of produced particles only.  We
make a remark that, if the initial state contains particles, we may
utilise \eref{eq:spectrum} to address the problem of particle
absorption as well as particle production.

Our goal at the moment is to find an alternative expression of
\eref{eq:spectrum} in terms of field variables instead of
creation/annihilation operators.  To this end we need a prescription
to identify creation/annihilation operators under background fields.
These operators are related to the field operator $\hatphi(x)$ via the
expansion on complete basis, which is a clean procedure in the
asymptotic states.  In the out-state the annihilation operator is
extracted through
\begin{eqnarray}
 \sqrt{2E_{\rm out}(\bp)}\,\hata_{\rm out}(\bp)
  &= \lim_{t\to\infty} \rmi\int \rmd^3 \bx\,
  \rme^{\rmi E_{\rm out}(\bp)t-\rmi\bp\cdot\bx}\,\bigl[\partial_t
  -\rmi E_{\rm out}(\bp)\bigr] \hatphi(t,\bx) \nonumber\\
 &= \lim_{t\to\infty} \rmi\, \rme^{\rmi E_{\rm out}(\bp)t}\,\bigl[\partial_t
  -\rmi E_{\rm out}(\bp)\bigr]\,\hatphi(t,\bp) \;.
\label{eq:a+}
\end{eqnarray}
We use the same notation $\hatphi$ also for the Fourier transformed
field as long as no confusion arises.  In our convention the
normalisation above is consistent with the commutation relation,
$[\hata_{\rm out}(\bp),\hatad_{\rm out}(\bp')]=(2\pi)^3\delta(\bp-\bp')$.
This \eref{eq:a+} is a basic relation frequently used in the
derivation of the LSZ reduction formula in many textbooks.  Because
the number operator involves the creation/annihilation operators at
the equal time, we can drop the exponential part and simplify the
formula as
\begin{equation}
\fl\quad
 \hatad_{\rm out}(\bp)\hata_{\rm out}(\bp) = \frac{1}{2E_{\rm out}(\bp)}
  \lim_{t_1=t_2=t\to\infty} \bigl[ \partial_{t_1}+\rmi E_{\rm out}(\bp) \bigr]\,
  \bigl[ \partial_{t_2}-\rmi E_{\rm out}(\bp) \bigr]\,
  \hatphi^\dag(t_1,\bp)\hatphi(t_2,\bp) \;.
\label{eq:ada}
\end{equation}
If we are interested in the production of anti-particles, we can find
a similar formula replacing $\hatad_{\rm out}(\bp)\hata_{\rm out}(\bp)$
with $\hatbd_{\rm out}(-\bp)\hatb_{\rm out}(-\bp)$.  Owing to the
conservation of U(1) charge (electric charge), the number of
anti-particles should be anyway identical with that of particles, so
we do not calculate it explicitly here.

In view of this form it is already clear that we can translate
\eref{eq:spectrum} into a representation by means of the Wightman
function \cite{Landsman:1986uw}.  It should be more illuminating to
find an alternative expression using the spectral functions instead of
the Wightman function, for the spectral functions provide us with more
intuition about physical contents of the system.  Let us define the
spectral functions or the Wigner transformed Pauli-Jordan's (denoted
by $\calA$) and Hadamard's (denoted by $\calD$) functions as follows;
\begin{eqnarray}
 \calA^\Delta(t,p_0,\bp) &:= \frac{1}{V}\int_{-\Delta}^\Delta \rmd\delt\,
  \rme^{\rmi p_0 \delt}\, \bigl\langle \hatrho_{\rm in}\,
  [\hatphi(t+\case{1}{2}\delt,\bp),\hatphi^\dag(t-\case{1}{2}\delt,\bp)
  ]\bigr\rangle \;,\nonumber\\
 \calD^\Delta(t,p_0,\bp) &:= \frac{1}{V}\int_{-\Delta}^\Delta \rmd\delt\,
  \rme^{\rmi p_0 \delt}\, \langle \hatrho_{\rm in}\,
  \{\hatphi(t+\case{1}{2}\delt,\bp),\hatphi^\dag(t-\case{1}{2}\delt,\bp)
  \}\rangle \;.
\label{eq:spect_def}
\end{eqnarray}
Here we put a volume factor $V$ because we look at the same momenta
$\bp$ and trivially there arises $2\pi\delta(0)=V$.  One could define
the spectral functions with two momentum arguments, which would be
useful in the presence of spatially modulated background fields.  In
this work, however, we consider only the spatially homogeneous case,
so that the above definition \eref{eq:spect_def} suffices for our
goal.  It should be mentioned that our definition of
\eref{eq:spect_def} explicitly depends on an extra parameter
$\Delta$.  In the Wigner transformation, usually, one formally takes
$\Delta\to\infty$.  For a practical application to the numerical
analysis, a finite $\Delta$ as above would be legitimate, as we will
discuss later.  To extract information on the in- or the out-state,
as a matter of fact, one should keep the ordering, $|t|\gg\Delta|$,
when we formally take $\Delta\to\infty$; otherwise the spectral
functions are affected by the interaction even for $t$ that is far
outside of the interacting region.  Roughly speaking, $\Delta$ should
be interpreted as an ``observation time'' with which the
quasi-particle oscillation is resolved.

We can change the variables from $t_1$ and $t_2$ to
$t=\case{1}{2}(t_1+t_2)$ and $\delt=t_1-t_2$, so that we can finally
arrive at the following formula,
\begin{equation}
 \fl\quad
 \frac{\rmd N}{\rmd^3\bp} = \lim_{t\to\infty} \frac{V}{(2\pi)^3}
  \int\frac{\rmd p_0}{2\pi}\,
  \frac{1}{4E_{\rm out}(\bp)} \Bigl[ \case{1}{4}\partial^2_t
  + \bigl(p_0+E_{\rm out}(\bp)\bigr)^2 \Bigr]
  \bigl[ \calD^\Delta(t,p_0,\bp) - \calA^\Delta(t,p_0,\bp) \bigr] \;.
\label{eq:spect}
\end{equation}
It is important to stress that \eref{eq:spect} does not rely on a
choice of the integration range $\Delta$ in the definition of
\eref{eq:spect_def}.  This is because the $p_0$-integration picks
$\delta(\delt)$ up to realize $t_1=t_2$ after taking each derivative
on $t_1$ and $t_2$.  Although the results should be the same
regardless of $\Delta$, the physical picture becomes more vivid if we
choose an appropriate value of $\Delta$.  Using that
$\calD^\Delta(t,p_0,\bp)$ and $\calA^\Delta(t,p_0,\bp)$ are even and
odd functions of $p_0$, respectively, we can readily confirm that the
contribution from anti-particles amounts to just the same answer,
which should be guaranteed by the charge conservation.

Here we note that our formula \eref{eq:spect} looks significantly
different from the LSZ-type expression as used in
\cite{Fukushima:2009er,Gelis:2013oca}.  We need go back to
\eref{eq:a+} and adopt \eref{eq:a+} as a \textit{definition} of the
annihilation operator at time $t$.  Then, we can pick
$\hata_{\rm out}(\bp)$ at $t=\infty$ up from the boundary if we
integrate the $t$-derivative of \eref{eq:a+} with respect to $t$.  The
$t$-derivative leads to $[\partial_t+\rmi E_{\rm out}(\bp)]$ on
$\hatphi(t,\bp)$, so we get $[\partial_t^2+E_{\rm out}^2(\bp)]$ as
usual in the LSZ reduction formula.  A problem in this argument is
that $E_{\rm out}(\bp)$ makes sense only at $t=\infty$ and there is no
clear-cut prescription to identify the dispersion relation for any
$t$.  In the adiabatic limit with slowly changing vector potential
$\bA(t)$, one may be able to approximate it as
$E(t,\bp)=\sqrt{[\bp+e\bA(t)]^2+m^2}$ as assumed in
\cite{Schmidt:1998zh,Bloch:1999eu} (see also \cite{Tanji:2008ku} for
more discussions on the physical interpretation).  Thus, with this
subtlety about the dispersion relation at intermediate time, we do not
think that the LSZ-type expression is any more advantageous than our
formula \eref{eq:spect}.

Only for completeness let us make a remark on another expression with
use of Green's functions.  We can perform the Wigner transform for the
retarded and advanced propagators to define
$D^\Delta_{\rm R}(t,p_0,\bp)$ and $D^\Delta_{\rm A}(t,p_0,\bp)$
as well as the Feynman (time-ordered) propagator,
$D^\Delta_{\rm F}(t,p_0,\bp)$.  Then, \eref{eq:spect} is just
equivalent with
\begin{equation}
 \fl\quad
 \frac{\rmd N}{\rmd^3\bp} = \lim_{t\to\infty}\frac{V}{(2\pi)^3}
  \int\frac{\rmd p_0}{2\pi}\,
  \frac{1}{2E_{\rm out}(\bp)} \Bigl[ \case{1}{4}\partial^2_t
  + \bigl(p_0+E_{\rm out}(\bp)\bigr)^2 \Bigr]
  \bigl[ D^\Delta_{\rm F}(t,p_0,\bp) - D^\Delta_{\rm R}(t,p_0,\bp)
  \bigr] \;.
\end{equation}
This expression might be more tractable if one wants to apply the
Schwinger-Keldysh formalism to compute the correlation functions.

To gain some feeling about how our formula works, we will take a quick
look at the typical behaviour of these spectral functions in the
asymptotic states where we can expand the field in terms of plane
waves.

\section{Spectral functions in the asymptotic states}

Because the dynamical spectral functions are less known objects than
more conventional ones in equilibrated matter, we will devote this
section to the exploration of how they look like in the asymptotic in-
and out-states.  To reduce unnecessary complication, we shall limit
our discussion to the choice of
$\hatrho_{\rm in}=|0_{\rm in}\rangle\langle 0_{\rm in}\rangle$ for the
moment.  We will address an extension to the finite
temperature/density environment in the later section.  We denote the
annihilation operators, $\hata_{\rm in}(\bp)$ and $\hatb_{\rm in}(\bp)$,
respectively, for particles and anti-particles, with which the vacuum
$|0_{\rm in}\rangle$ is defined.  For a practical purpose we take the
range of $t$ from $-T$ to $T$ with a sufficiently large $T$.

The field operator in the in-state around $t=-T$ is then a
superposition of the plane waves with $\hata_{\rm in}(\bp)$ for the
positive-energy oscillation and $\hatb^\dag_{\rm in}(\bp)$ for the
negative-energy oscillation, i.e.
\begin{equation}
 \hatphi(t\sim-T,\bp) = \frac{1}{\sqrt{2E_{\rm in}(\bp)}}\Bigl[
  \hata_{\rm in}(\bp)\,\rme^{-\rmi E_{\rm in}(\bp) t}
  +\hatbd_{\rm in}(-\bp)\,\rme^{\rmi E_{\rm in}(\bp) t} \Bigr] \;.
\label{eq:initial}
\end{equation}
The plane waves should be the solutions of the equation of motion
around $t=-T$, and they evolve to a mixture of the positive- and the
negative-energy states as $t$ elapses toward the interacting region.
We can then parametrise this mixing effect as
\begin{equation}
 \frac{\rme^{-\rmi E_{\rm in}(\bp)t}}{\sqrt{2E_{\rm in}(\bp)}}
 \;\;(t\sim-T)
 \;\longrightarrow\;
 \frac{\alpha_{\bp}\,\rme^{-\rmi E_{\rm out}(\bp)t} + \beta^\ast_{\bp}\,
  \rme^{\rmi E_{\rm out}(\bp)t}}{\sqrt{2E_{\rm out}(\bp)}} \;\;(t\sim T) \;,
\end{equation}
where the Bogoliubov coefficients, $\alpha_{\bp}$ and $\beta_{\bp}$,
are determined according to the equation of motion, and a similar
relation should hold for another branch of solution starting with
$\propto \rme^{\rmi E_{\rm in}(\bp)t}$.  In fact, if the Hamiltonian is
Hermite, the complex conjugate of the above relation is true, so the
field operator in the out-state at $t=T$ then reads,
\begin{eqnarray}
 \hatphi(t\sim\infty,\bp) = \frac{1}{\sqrt{2E_{\rm out}(\bp)}}\Bigl\{
 & \bigl[ \alpha_{\bp}\hata_{\rm in}(\bp)+\beta_{\bp}\hatbd_{\rm in}(-\bp)
  \bigr] \rme^{-\rmi E_{\rm out}(\bp)t} \nonumber\\
 & \qquad +\bigl[ \alpha^\ast_{\bp}\hatbd_{\rm in}(-\bp)
        +\beta^\ast_{\bp}\hata_{\rm in}(\bp) \bigr]
  \rme^{\rmi E_{\rm out}(\bp)t} \Bigr\} \;.
\label{eq:final}
\end{eqnarray}
This defines the creation/annihilation operators in the out-state, and
by requiring the canonical commutation relation for them, we can find
the normalisation condition, $|\alpha_{\bp}|^2 - |\beta_{\bp}|^2 = 1$.

At this point we can immediately recover the known result for the
Schwinger mechanism directly from \eref{eq:ada}.  Applying the
operator $[\partial_t-\rmi E_{\rm out}(\bp)]$ on $\hatphi$ we project
the positive-energy part out, and then we can plug the number operator
of \eref{eq:ada} into \eref{eq:spectrum}, which yields an estimate of
produced particles as
\begin{equation}
 \frac{\rmd N}{\rmd^3\bp} = \frac{1}{(2\pi)^3}
  \langle 0_{\rm in}| \beta^\ast_{\bp}\hatb_{\rm in}(-\bp) \,
  \beta_{\bp}\hatbd_{\rm in}(-\bp) |0_{\rm in}\rangle
 = \frac{V|\beta_{\bp}|^2}{(2\pi)^3} \;.
\label{eq:answer}
\end{equation}
This is a standard formula for the particle production obtained via
the Bogoliubov transformation \cite{Nikishov:1970br}.  Now it is
intriguing to check how our formula \eref{eq:spect} gives rise to the
same answer.

We can immediately compute the spectral functions from the asymptotic
forms \eref{eq:initial} and \eref{eq:final} if we take $T\gg\Delta$.
Then, the spectral functions at $t=-T$ have no access to the region
with non-vanishing background fields, so they take a familiar
expression just for non-interacting particles;
\begin{eqnarray}
 \calA^\Delta(t\sim-T,p_0,\bp) = \frac{\pi}{E_{\rm in}(\bp)}\Bigl[
  \delta(p_0-E_{\rm in}(\bp)) - \delta(p_0+E_{\rm in}(\bp)) \Bigr]\;,\\
 \calD^\Delta(t\sim-T,p_0,\bp) = \frac{\pi}{E_{\rm in}(\bp)}\Bigl[
  \delta(p_0-E_{\rm in}(\bp)) + \delta(p_0+E_{\rm in}(\bp)) \Bigr]\;.
\label{eq:DAin}
\end{eqnarray}
In this case $\calD^\Delta-\calA^\Delta$ has only a term that is
proportional to $\delta(p_0+E_{\rm in}(\bp))$, and thus the produced
particle is vanishing as is obvious from \eref{eq:spect}.  Now let us
go into later time when these functions should change their shape.
Once \eref{eq:final} eventually follows, it is just a simple
arithmetic procedure to reach,
\begin{eqnarray}
\fl\quad \calA^\Delta(t\sim T,p_0,\bp) &= \frac{\pi}
  {E_{\rm out}(\bp)}\Bigl[\delta(p_0-E_{\rm out}(\bp))
  - \delta(p_0+E_{\rm out}(\bp)) \Bigr]\;,\\
\fl\quad \calD^\Delta(t\sim T,p_0,\bp) &= \Bigl[
  |\alpha_{\bp}|^2+|\beta_{\bp}|^2 \Bigr]\,\frac{\pi}{E_{\rm out}(\bp)}
  \,\Bigl[ \delta(p_0-E_{\rm out}(\bp)) + \delta(p_0+E_{\rm out}(\bp))
  \Bigr] \nonumber\\
 &\qquad\qquad\qquad\qquad\qquad
  + \frac{2}{E_{\rm out}(\bp)}\mathrm{Re}\Bigl[ \alpha_{\bp}\beta_{\bp}\,
  \rme^{-2\rmi E_{\rm out}(\bp)t} \Bigr]\,2\pi\delta(p_0) \;.
\label{eq:Dout}
\end{eqnarray}
Here, again, we required $T\gg\Delta$.  There are two interesting
observations as perceived from the above:  (1) Pauli-Jordan's function
$\calA^\Delta$ is insensitive to the Bogoliubov transformation and the
overall factor is $|\alpha_{\bp}|^2-|\beta_{\bp}|^2=1$.  (2)
Hadamard's function $\calD^\Delta$ is affected by the mixing effect by
$|\alpha_{\bp}|^2+|\beta_{\bp}|^2\neq1$ and, besides, an interference
term $\propto\alpha_{\bp}\beta_{\bp}$ appears.  We emphasise that such
an interference term is usually absent and is quite peculiar to the
Bogoliubov mixing effect.

Then, the difference between these two spectral functions consists of
three terms as follows,
\begin{eqnarray}
 \calD^\Delta - \calA^\Delta &= |\beta_{\bp}|^2 \frac{2\pi}
  {E_{\rm out}(\bp)} \delta(p_0 - E_{\rm out}(\bp)) + |\alpha_{\bp}|^2
  \frac{2\pi}{E_{\rm out}(\bp)} \delta(p_0 + E_{\rm out}(\bp)) \nonumber\\
 &\qquad\qquad\qquad\qquad + \frac{2}{E_{\rm out}(\bp)}\mathrm{Re}
  \bigl[ \alpha_{\bp}\beta_{\bp} \rme^{-2\rmi E_{\rm out}(\bp)t}\bigr]
  2\pi\delta(p_0) \,.
\label{eq:dafinal}
\end{eqnarray}
We can make it sure by the explicit calculation that the second
($\propto |\alpha_{\bp}|^2$) and the third
($\propto \alpha_{\bp}\beta_{\bp}$) terms have no finite contribution
if applied to \eref{eq:spect}, and only the first
($\propto |\beta_{\bp}|^2$) term is relevant to the particle
production, which yields exactly the same answer as \eref{eq:answer}.

Although the calculations are very easy, the expression of
\eref{eq:dafinal} in the out-state has profound implications.  In many
situations we typically have $\alpha_{\bp}\approx 1$ and
$|\beta_{\bp}|\ll 1$, for which the first term is much smaller than
the third interference term.  In the next section, indeed, we will
numerically compute the spectral functions and confirm that this is
the case.  It would be an interesting future problem to think of a way
to make use of the interference term in order to probe the Bogoliubov
mixing effect experimentally.

Before closing this section, it would be instructive to understand how
the standard propagators are modified by the Bogoliubov
transformation.  Surprisingly, we find that the retarded propagator is
intact under the mixing effect and only the Feynman propagator depends
on the Bogoliubov coefficients.  That is,
\begin{eqnarray}
 &\fl D^\Delta_{\rm R}(t\sim T,p_0,\bp) =
  \mathrm{P}\frac{\rmi}{p_0^2-E_{\rm out}^2(\bp)}
  + \frac{\pi}{2E_{\rm out}(\bp)}
  \Bigl[ \delta(p_0-E_{\rm out}(\bp)) - \delta(p_0+E_{\rm out}(\bp))
    \Bigr] \;,\\
 &\fl D^\Delta_{\rm F}(t\sim T,p_0,\bp) =
  \mathrm{P}\frac{\rmi}{p_0^2-E_{\rm out}^2(\bp)} + \bigl(
  |\alpha_{\bp}|^2 \!+\! |\beta_{\bp}|^2 \bigr)
  \frac{\pi}{2E_{\rm out}(\bp)} \Bigl[ \delta(p_0\!-\!E_{\rm out}(\bp))
  + \delta(p_0\!+\!E_{\rm out}(\bp)) \Bigr] \nonumber\\
 &\qquad\qquad\qquad\qquad\qquad\qquad\qquad
 + \frac{1}{E_{\rm out}(\bp)}\mathrm{Re}\bigl(\alpha_{\bp}\beta_{\bp}
  \rme^{-2\rmi E_{\rm out}(\bp)t}\bigr) 2\pi\delta(p_0) \;,
\end{eqnarray}
where P stands for taking the principal value.  It is clear at glance
that $\calD^\Delta-\calA^\Delta=2(D^\Delta_{\rm F}-D^\Delta_{\rm R})$
holds as it should.

Now it is time to take one step forward to understand how the spectral
functions should evolve continuously from the in-state to the
out-state as a function of $t$.  In the aim of visualising the
behaviour with increasing $t$, we need to perform numerical
calculations.  In the next section we present our numerical results.

\section{Spectral functions in pulsed electric fields}

We here solve the equation of motion for given electric fields.  In
principle, we can numerically deal with arbitrary electric fields
within our present approximation to neglect the back-reaction.
Although the analytical solution is not demanded here, we shall choose
one of the most well-investigated profile known as the Sauter
potential \cite{Narozhnyi:1970uv}, which is solvable and identifiable
with a pulsed electric field,
\begin{equation}
 E(t) = E\,\mathrm{sech}^2(\omega t) \;.
\end{equation}
The frequency parameter $\omega$ characterises the life time of the
applied electric field.  Let us choose the $z$ axis along the
direction of the electric field, and then the associated vector
potential reads,
\begin{equation}
 A_z(x) = \frac{E}{\omega}\bigl[ \tanh(\omega t)-1 \bigr] \;.
\end{equation}
Then, we can find two independent solutions, $\psi_p^{(\pm)}(t)$, by
solving the equation of motion under this vector potential,
\begin{equation}
 \Bigl[ \partial_t^2 + \Bigl( p_z+\frac{eE}{\omega} \bigl[\tanh(\omega t)
  -1\bigr]\Bigr)^2 + m_\perp^2 \Bigr] \psi_p^{(\pm)}(t) = 0 \;,
\label{eq:eom}
\end{equation}
where $m_\perp$ represents the transverse mass,
$m_\perp^2:=p_x^2+p_y^2+m^2$.  We should impose the following boundary
conditions;
\begin{equation}
 \psi_p^{(\pm)}(t=-T) = \frac{1}{\sqrt{2E_{\rm in}(\bp)}}\,
  \rme^{\mp\rmi E_{\rm in}(\bp) t} \;,
\label{eq:boundary}
\end{equation}
for large enough $T$, with the dispersion relations,
\begin{equation}
 E_{\rm in}(\bp) = \sqrt{(p_z-2eE/\omega)^2+m_\perp^2}\;,\qquad
 E_{\rm out}(\bp) = \sqrt{p_z^2+m_\perp^2}\;.
\end{equation}

\begin{figure}
\begin{center}
 \includegraphics[width=0.72\textwidth]{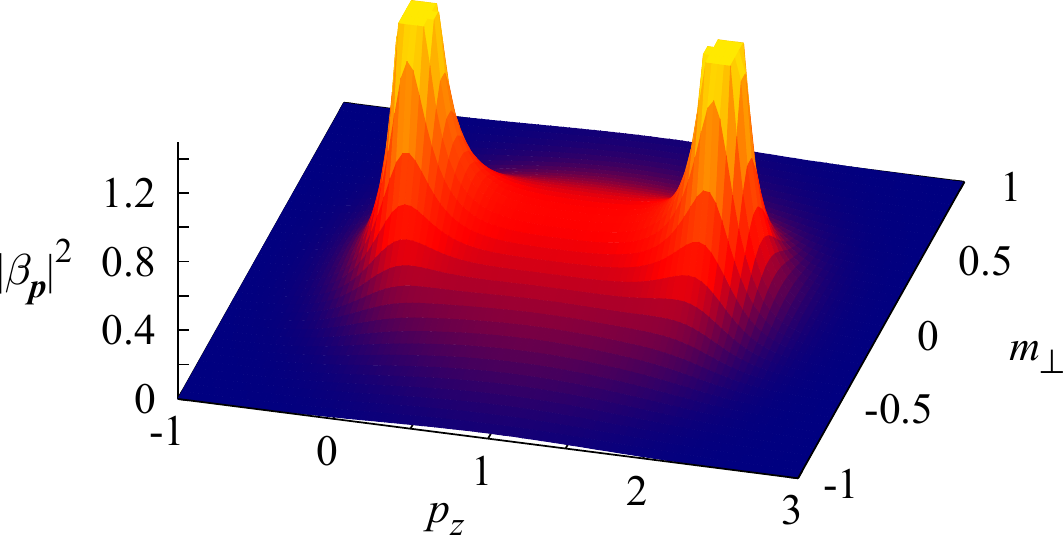}
 \caption{Bogoliubov coefficient $|\beta_{\bp}|^2$ for the choice of
   $\omega/\sqrt{eE}=1$.  All quantities are measured in unit of
   $\sqrt{eE}$.}
 \label{fig:analytic}
\end{center}
\end{figure}

One can write the analytical expressions of $\psi_p^{(\pm)}(t)$ using
the hyper-geometric functions.  Therefore, the number of produced
particle or $|\beta_{\bp}|^2$ is analytically known.  Hereafter we
shall refer to all quantities with mass dimensions in unit of the
electric field $\sqrt{eE}$ and present our results with dimensionless
numbers.  In this work we work with a specific choice of
\begin{equation}
 \frac{\omega}{\sqrt{eE}}=1\;,
\label{eq:omega}
\end{equation}
to investigate the effect of pulsed electric fields.  A different
choice of $\omega$ makes no qualitative change in our resulting
spectral functions.  We make a plot in \fref{fig:analytic} to show the
analytical structure of $|\beta_{\bp}|^2$ as a function of the
longitudinal momentum $p_z$ and the transverse mass $m_\perp$.

Obviously from \fref{fig:analytic}, the particle number becomes
greater for smaller $m_\perp$.  The produced particles are
accelerated to the positive $z$ direction by the electric field, and
as understood from \eref{eq:eom}, $p_z$ is shifted by
$0\sim 2eE/\omega$ during the time evolution.  This means that the
momentum distribution of the produced particles should spread over
$p_z=0\sim 2eE/\omega$.  We can confirm this expectation explicitly on
\fref{fig:analytic}.

To discuss the effect of the particle production in a reasonably
visible manner, we will look at the point of following momenta,
\begin{equation}
 \frac{m_\perp}{\sqrt{eE}}=0\;,\qquad
 \frac{p_z}{\sqrt{eE}}=\frac{\sqrt{eE}}{\omega}=1.5\;,
\label{eq:pz}
\end{equation}
which deviates from the pronounced peak seen in \fref{fig:analytic}.
One might have thought that the exact peak position would be a better
choice, but if we choose $m_\perp=0$ and $p_z=2$, for instance, the
numerical calculations result in singularity out of control.

\begin{figure}
\begin{center}
 \includegraphics[width=0.6\textwidth]{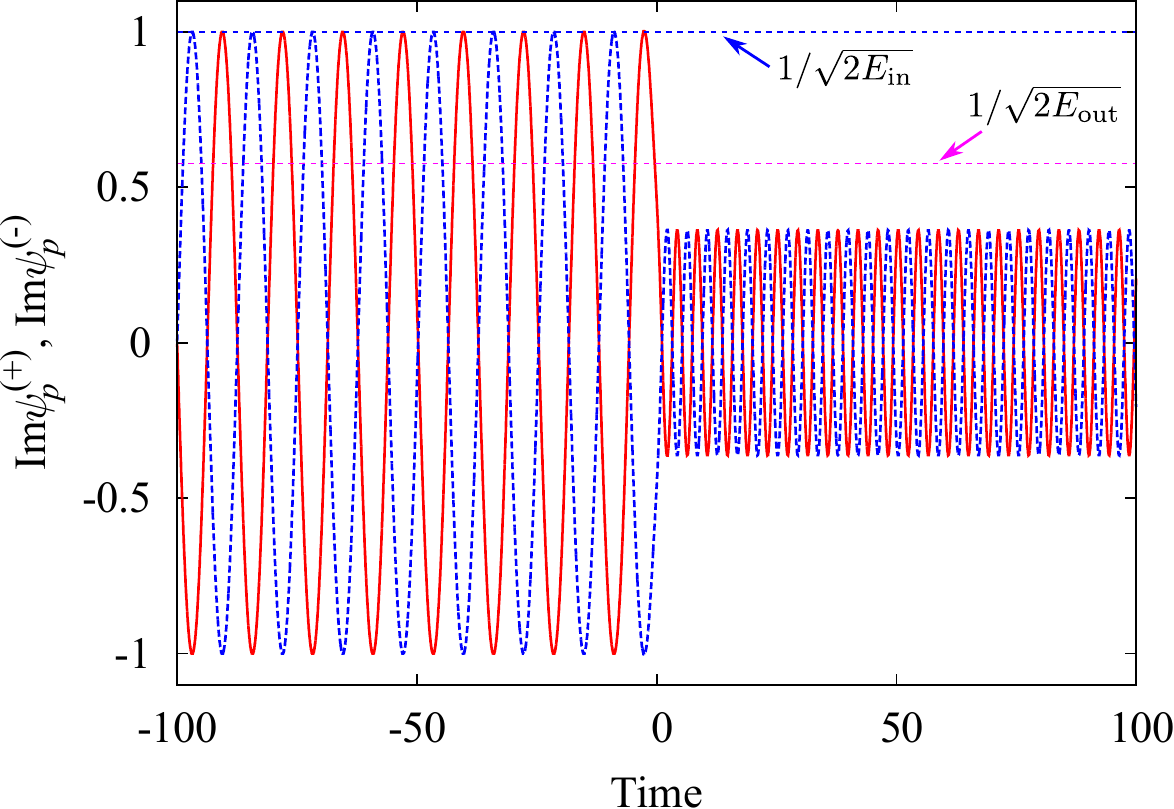}
 \caption{Two independent wave functions satisfying the given
   boundary conditions \eref{eq:boundary} at $t=-T$ with $T=100$.}
 \label{fig:sol}
\end{center}
\end{figure}

With these parameters we solve the equation of motion \eref{eq:eom}
numerically to find $\psi_p^{(\pm)}(t)$, the results of which are
shown in \fref{fig:sol}.  We used the 4th-order Runge-Kutta (RK4)
method and took 20000 points to discretise along the time direction.
We imposed the boundary conditions \eref{eq:boundary} at $t=-T$ with
$T=100$. Because the equation of motion is real,
$\psi_p^{(-)}(t)=\psi_p^{(+)*}(t)$ follows immediately, and this means
that the real part of them should be identical.  This is why we
present the imaginary part in \fref{fig:sol}, and indeed, we can make
it sure that our numerical calculations go correctly to respect
${\rm Im}\psi_p^{(-)}(t)=-{\rm Im}\psi_p^{(+)}(t)$.

The in-state around $t\sim-T$ has the field amplitude of the correct
normalisation $1/\sqrt{2E_{\rm in}(\bp)}$, while at later time, as
seen in \fref{fig:sol}, the amplitude deviates from
$1/\sqrt{2E_{\rm out}(\bp)}$.  This discrepancy is attributed to the
mixing between the positive- and the negative-energy states and thus
signals for the Bogoliubov transformation.

\begin{figure}
\begin{center}
 \includegraphics[width=0.6\textwidth]{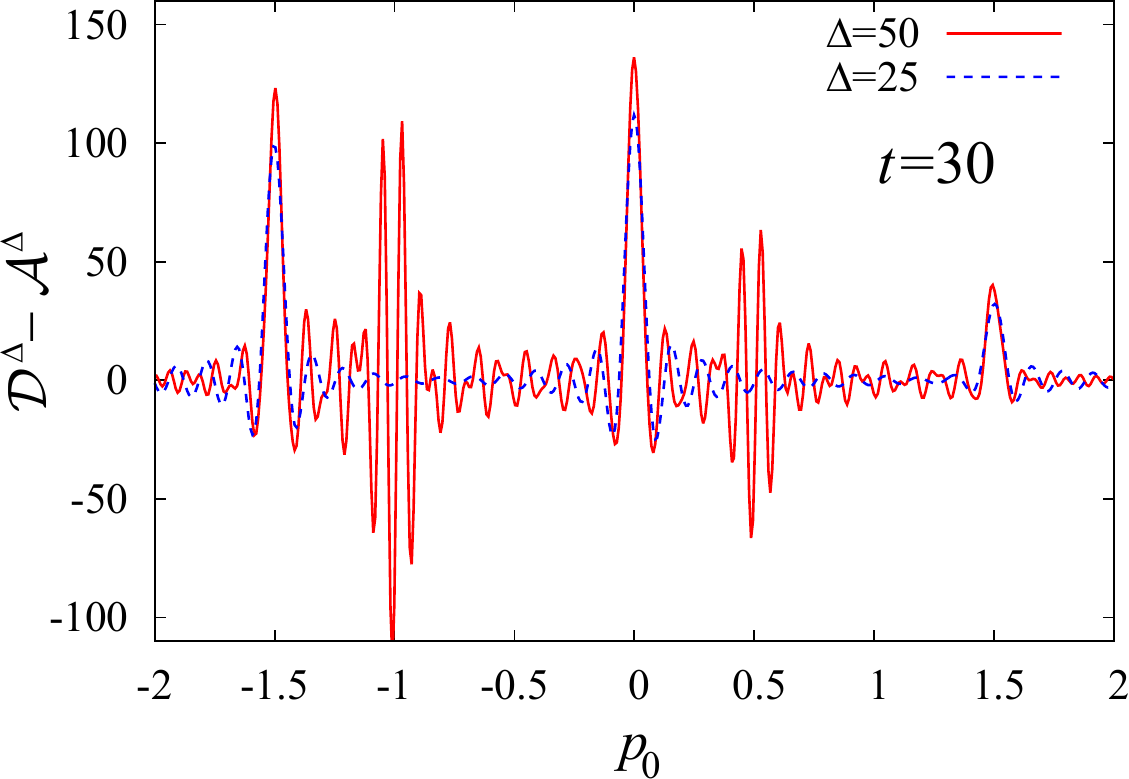}
 \caption{$\Delta$ dependence of the spectral function
  $\calD^\Delta-\calA^\Delta$ at $t=30$.  The solid (and dashed) curve
   represents the results at $\Delta=50$ (and $25$, respectively).}
 \label{fig:delta}
\end{center}
\end{figure}

Once we have the wave-functions, we can construct the spectral
functions for any $t$, i.e.\ a simple calculation leads to
\begin{equation}
\fl\quad
 \calA^\Delta(t,p_0,\bp) = \int_{-\Delta}^\Delta \rmd\delt\,
  \rme^{\rmi p_0 \delt} \biggl[ \psi_{\bp}^{(+)}(t+\case{1}{2}\delt)
  \psi_{\bp}^{(+)\ast}(t-\case{1}{2}\delt)
  - \psi_{\bp}^{(-)}(t+\case{1}{2}\delt)
  \psi_{\bp}^{(-)\ast}(t-\case{1}{2}\delt) \biggr] ,
\label{eq:spectA}
\end{equation}
for Pauli-Jordan's function and we can find a similar expression for
Hadamard's function.  The time evolution of the spectral functions may
have dependence on the choice of $\Delta$.  Intuitively, $\Delta$
corresponds to the observation time, as we already mentioned, to
detect the quasi-particle behaviour in the oscillation pattern.  For
the concrete demonstration, let us take a look at \fref{fig:sol}
again; the temporal oscillation shows a constant pattern except near
the origin where the system is disturbed by pulsed electric fields.
So, around $t=30$ for example, if $\Delta$ is less than $30$, the
quasi-particle behaviour is well separated from the interaction region
at the origin and the spectral functions should be close to
\eref{eq:dafinal} then.  If $\Delta$ is greater then $30$, however,
the integration region covers the pulsed electric fields, which should
alter the spectral shape.  Indeed, as we can see in \fref{fig:delta},
we can confirm this anticipation by comparing the results at
$\Delta=25< t=30$ and $\Delta=50$ for $\calD^\Delta-\calA^\Delta$.

\Fref{fig:delta} already indicates the Schwinger process of the
particle production.  We can see a peak at
$p_0\sim E_{\rm out}(\bp)=1.5$ and its height corresponds to
$|\beta_{\bp}|^2$ according to \eref{eq:spect}.  Precisely speaking,
if we take $\Delta\to\infty$, the peak becomes a Dirac's delta
function and the $p_0$-integration in \eref{eq:spect} has a
contribution from a point $p_0=E_{\rm out}(\bp)$ only.  Now that we
implement the Wigner transformation with a finite $\Delta$, the peak
is broadened and we should keep the $p_0$-integration over the range
of the order of $1/\Delta$.

From \eref{eq:dafinal} we can understand two more peaks are expected
at $p_0\sim -E_{\rm out}(\bp)=-1.5$ and $p_0\sim 0$ which are evident in
\fref{fig:delta}.  For $\Delta=50$, we are also aware of some
enhancement around $p_0\sim -1$ and $0.5$, which is quite non-trivial.
They arise from the effect of the background fields when $\Delta$ is
comparable to or greater than $t$.

\begin{figure}
\begin{center}
 \includegraphics[width=0.8\textwidth]{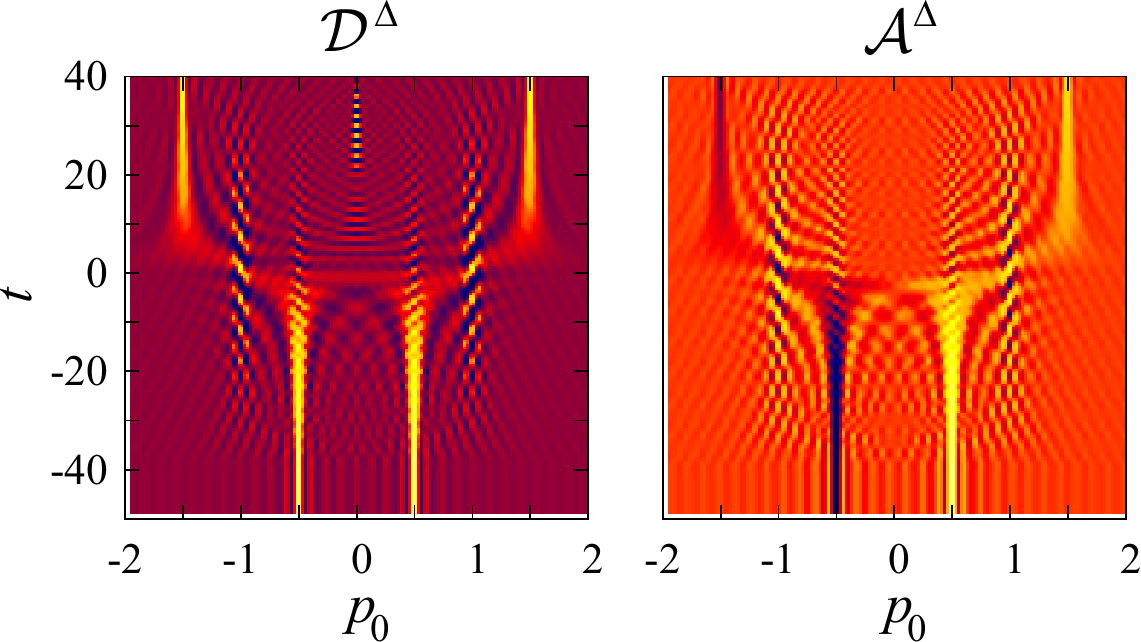}
 \caption{Density plots of the spectral functions,
 $\calD^\Delta$ and $\calA^\Delta$, for the choice of
   \eref{eq:omega} and \eref{eq:pz} and $\Delta=40$.  The bright
   (and dark) colour indicates a larger (and smaller, respectively)
   value.}
 \label{fig:spectDA}
\end{center}
\end{figure}

\begin{figure}
\begin{center}
 \includegraphics[width=0.72\textwidth]{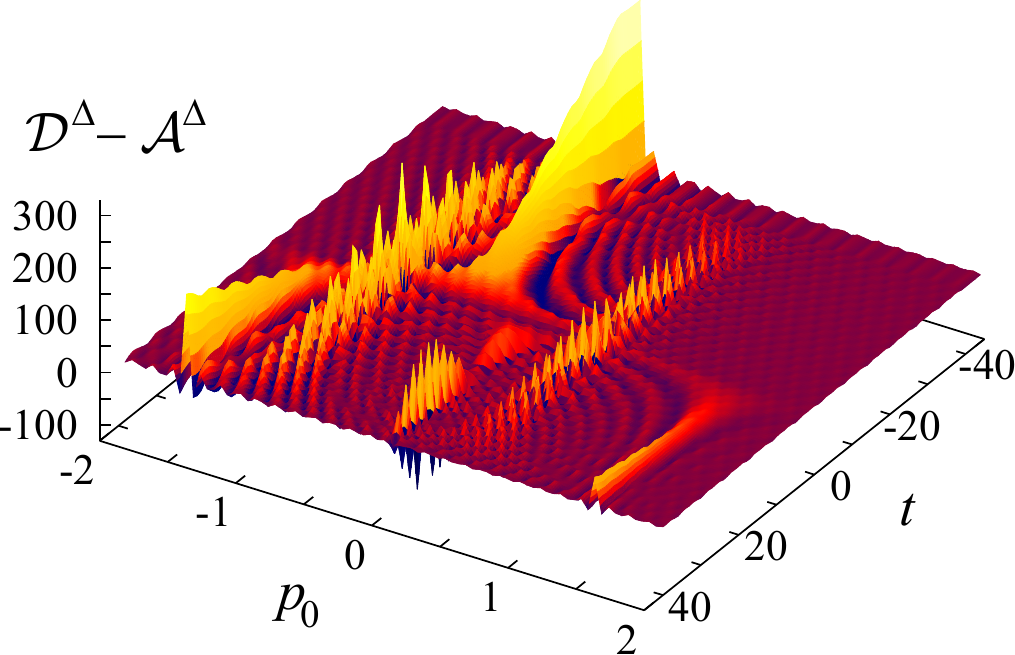}
 \caption{Evolution of the spectral function difference
   $\calD^\Delta-\calA^\Delta$ for the choice of \eref{eq:omega} and
   \eref{eq:pz} with $\Delta=40$.}
 \label{fig:spect}
\end{center}
\end{figure}

We are interested in the full temporal profile of these enhanced
regions, so we make a density plot for $\calD^\Delta$ and
$\calA^\Delta$ individually, as is shown in \fref{fig:spectDA}.  This
figure provides us with useful messages about the flow of the spectral
peaks.  First of all, the oscillatory peak at $p_0=0$ appears only in
$\calD^\Delta$ at late $t$, as is the case in the out-state
\eref{eq:Dout}.  Second, we can notice that the intermediate
enhancement around $p_0\sim -1$ and $0.5$ emerges in both
$\calD^\Delta$ and $\calA^\Delta$.  Because the enhancement originates
from time-dependent background fields, it would be conceivable that no
simple pattern but complicated time dependence may well occur.  This
is not the case, however, and the enhancement goes rather straight in
time.  We can observe this in a clearer way in the form of not the
density plot but the three-dimensional (3D) plot as presented in
\fref{fig:spect}.

In \fref{fig:spect} we plot $\calD^\Delta-\calA^\Delta$ that is the
difference between two in \fref{fig:spectDA}.  Near the in-state, as
seen in \fref{fig:spect}, there stands only one peak in the vicinity
of $p_0=-E_{\rm in}(\bp)$, which agrees perfectly with the asymptotic
analysis \eref{eq:DAin}.  Then, this peak diminishes with increasing
time, and meanwhile, the intermediate oscillatory modes grow up at
$p_0\sim -1$ and $0.5$.  Eventually, these modes fade away, and at the
same time, the spectral function approaches the asymptotic form of
\eref{eq:dafinal}.  At late time we can recognise a small peak around
$p_0\sim 1.5$ and this peak amounts to the Schwinger mechanism.  In
other words, the Schwinger mechanism is a phenomenon that takes
account of such a small portion of the whole spectral shape.  It is
certainly worth considering other tools to diagnose a wider region of
the spectral functions, which is an interesting future problem beyond
the present scope.

\section{Extension to the finite temperature}

Finally we shall extend our analysis to a more general situation of the
initial state.  A more non-trivial but still controllable example is
the finite temperature/density calculation of the Schwinger mechanism
\cite{Kim:2008em}.  Let us assume that the initial state is a mixed
state characterised by the following density matrix,
\begin{equation}
 \hat{\rho}_\infty = \frac{\exp\bigl[ -\beta E_{\rm in}(\bp)\,
  \hata^\dag_{\rm in}(\bp) \hata_{\rm in}(\bp) \bigr]}{\mathrm{tr}\bigl\{{
  \exp\bigl[ -\beta E_{\rm in}(\bp)\, \hata^\dag_{\rm in}(\bp)
  \hata_{\rm in}(\bp) \bigr]\bigr\}}} \;,
\end{equation}
with $\beta$ being the inverse temperature.  Then the straightforward
calculation immediately leads to the produced particle number given as
\begin{equation}
 \frac{\rmd N}{\rmd^3\bp} - \frac{V f_{\bp}}{(2\pi)^3}
 = \frac{V|\beta_{\bp}|^2}{(2\pi)^3}
   \Bigl( 1 + f_{\bp} + \bar{f}_{-\bp} \Bigr)\;,
\label{eq:nT}
\end{equation}
where $f_{\bp}$ and $\bar{f}_{\bp}$ represent the Bose-Einstein
distribution function, respectively, for the particle and the
anti-particle with the in-state energy $E_{\rm in}(\bp)$, namely,
\begin{equation}
 f_{\bp}:= \frac{1}{\rme^{\beta[E_{\rm in}(\bp)-\mu]}-1} \;,\qquad
 \bar{f}_{\bp}:= \frac{1}{\rme^{\beta[E_{\rm in}(\bp)+\mu]}-1}
\end{equation}
with a chemical potential $\mu$ introduced.  We note that, in the
left-hand side of \eref{eq:nT}, the number of thermal particles is
subtracted since they are irrelevant to the particle production.  This
result is understandable also from the spectral functions.  We can
find that Hadamard's function picks the Bose-Einstein distribution
function up as
\begin{eqnarray}
 \calD^\Delta(t,p_0,\bp) &= \int_{-\Delta}^\Delta \rmd\delt\,
  \rme^{\rmi p_0 \delt} \biggl[ \psi_{\bp}^{(+)}(t+\case{1}{2}\delt)
  \psi_{\bp}^{(+)\ast}(t-\case{1}{2}\delt) (1+2f_{\bp}) \nonumber\\
  &\qquad\qquad\qquad + \psi_{\bp}^{(-)}(t+\case{1}{2}\delt)
  \psi_{\bp}^{(-)\ast}(t-\case{1}{2}\delt) (1+2\bar{f}_{-\bp}) \biggr] ,
\end{eqnarray}
but Pauli-Jordan's function $\calA^\Delta$ remains independent of the
medium effect and is not changed from the vacuum expression
\eref{eq:spectA}.  Such a qualitative difference between
$\calD^\Delta$ and $\calA^\Delta$ makes a sharp contract and is not
quite comprehensible on the intuitive level.

The particle production is increased by the Bose-Einstein distribution
function.  Therefore, if $f_{\bp}$ takes a macroscopic value, this
increase must be a sizable effect.  In the high-$T$ limit, in fact,
the distribution function approaches,
$f_{\bp}\sim\bar{f}_{\bp}\to T/E_{\rm in}(\bp)$, and the particle
production is significantly enhanced by $2T/E_{\rm in}(\bp)$, which is
a substantial factor if $T$ is large or $E_{\rm in}(\bp)$ is small
enough, as is the case in the quark-gluon plasma.  However, the
in-state already contains as many particles as $f_{\bp}$ and so the
particle production is not practically enhanced if measured relative
to the number of particles in the in-state.

Another interesting limit lies in a finite chemical potential that
makes $\bar{f}_{\bp}$ (or $f_{\bp}$) be much bigger than $f_{\bp}$ (or
$\bar{f}_{\bp}$).  This may well opens a new possibility for the
experimental detection of the Schwinger process.  For example, if we
have a macroscopic occupation number like the Bose-Einstein condensate
of scalar particles (that is actually a superconductor), the
anti-particle (hole) that did not exist in the in-state is produced
with a gigantic enhancement factor by the macroscopic occupation
number.  It may be worth pursuing this possibility further in the
future research.

\section{Conclusions}

We found a useful formula that relates the particle production to the
dynamical (time-dependent) spectral functions.  We then clarified the
basic properties of these spectral functions and proceeded to the
numerical calculation of the spectral functions using the solutions of
the equation of motion for a complex scalar field theory under pulsed
electric fields.  We closely studied the time evolution of the
spectral functions.  Wigner transformed Hadamard's function turned out
to exhibit an oscillatory mode at $p_0=0$ as a result of the
Bogoliubov mixing.  This peak is larger by one power of the Bogoliubov
coefficient as compared to the other peak corresponding to the
Schwinger mechanism.  This structure hints a new possibility of
measurement that verifies the Bogoliubov mixing.  Another non-trivial
finding is the appearance of transient enhancement in the intermediate
time region.  In spite of time-dependent background fields, the
enhancement occurs somehow in an organised manner, which implies that
some unknown mechanism underlies the real-time dynamics.  We also
extended our discussions to the finite temperature/density case to
identify the medium enhancement factor.

Apart from the Schwinger problem, from a more general perspective of
theoretical physics, the dynamical properties of the spectral
functions are quite non-trivial and are still less unknown than the
equilibrated matter.  We emphasise that these spectral functions are
essential ingredients to think of real-time physics, and the particle
production is actually one of the possible applications.  In this
sense we should continuously invest our efforts to deepen the
theoretical understanding of the dynamical spectral functions and the
present work should contribute to the first step along this
direction.

\ack
This work was supported by JSPS KAKENHI Grant Number 24740169.

\section*{References}
\bibliographystyle{iopart-num}
\bibliography{schwinger}

\providecommand{\newblock}{}
\begin{thebibliography}{10}
\expandafter\ifx\csname url\endcsname\relax
  \def\url#1{{\tt #1}}\fi
\expandafter\ifx\csname urlprefix\endcsname\relax\def\urlprefix{URL }\fi
\providecommand{\eprint}[2][]{\url{#2}}

\bibitem{Schwinger:1960qe}
Schwinger J~S 1961 {\em J. Math. Phys.\/} {\bf 2} 407--432

\bibitem{Keldysh:1964ud}
Keldysh L 1964 {\em Zh. Eksp. Teor. Fiz.\/} {\bf 47} 1515--1527

\bibitem{kadanoff1962quantum}
Kadanoff L and Baym G 1962 {\em Quantum statistical mechanics: Green's function
  methods in equilibrium and nonequilibrium problems\/} Frontiers in physics
  (W.A. Benjamin)

\bibitem{Ivanov:2000}
Ivanov Y, Knoll J and Voskresensky D 2000 {\em Nucl. Phys.\/} {\bf A672}
  313--356 (\textit{Preprint} \eprint{nucl-th/9905028})

\bibitem{Kita:2006}
Kita T 2006 {\em J. Phys. Soc. Jpn.\/} {\bf 75} 114005

\bibitem{bellac1996thermal}
Bellac M 1996 {\em Thermal Field Theory\/} Cambridge Monographs on Mathematical
  Physics (Cambridge University Press) ISBN 9780521460408

\bibitem{abrikosov2012methods}
Abrikosov A, Gorkov L, Dzyaloshinski I and Silverman R 2012 {\em Methods of
  Quantum Field Theory in Statistical Physics\/} Dover Books on Physics (Dover
  Publications) ISBN 9780486140155

\bibitem{McLerran:1984ay}
McLerran L~D and Toimela T 1985 {\em Phys. Rev.\/} {\bf D31} 545

\bibitem{Weldon:1990iw}
Weldon H 1990 {\em Phys. Rev.\/} {\bf D42} 2384--2387

\bibitem{Heisenberg:1935qt}
Heisenberg W and Euler H 1936 {\em Z. Phys.\/} {\bf 98} 714--732
  (\textit{Preprint} \eprint{physics/0605038})

\bibitem{Schwinger:1951nm}
Schwinger J~S 1951 {\em Phys. Rev.\/} {\bf 82} 664--679

\bibitem{Dunne:2004nc}
Dunne G~V 2004  (\textit{Preprint} \eprint{hep-th/0406216})

\bibitem{Aoki:2013}
Aoki H, Tsuji N, Eckstein M, Kollar M, Oka T and Werner P 2013
  (\textit{Preprint} \eprint{1310.5329})

\bibitem{Soffel:1982pm}
Soffel M, Muller B and Greiner W 1982 {\em Phys. Rept.\/} {\bf 85} 51--122

\bibitem{Parikh:1999mf}
Parikh M~K and Wilczek F 2000 {\em Phys. Rev. Lett.\/} {\bf 85} 5042--5045
  (\textit{Preprint} \eprint{hep-th/9907001})

\bibitem{Fukushima:2009er}
Fukushima K, Gelis F and Lappi T 2009 {\em Nucl. Phys.\/} {\bf A831} 184--214
  (\textit{Preprint} \eprint{0907.4793})

\bibitem{Gelis:2006yv}
Gelis F and Venugopalan R 2006 {\em Nucl. Phys.\/} {\bf A776} 135--171
  (\textit{Preprint} \eprint{hep-ph/0601209})

\bibitem{Gelis:2006cr}
Gelis F and Venugopalan R 2006 {\em Nucl. Phys.\/} {\bf A779} 177--196
  (\textit{Preprint} \eprint{hep-ph/0605246})

\bibitem{Gelis:2013oca}
Gelis F and Tanji N 2013 {\em Phys. Rev.\/} {\bf D87} 125035 (\textit{Preprint}
  \eprint{1303.4633})

\bibitem{Berges:2012ks}
Berges J, Blaizot J~P and Gelis F 2012 {\em J. Phys.\/} {\bf G39} 085115
  (\textit{Preprint} \eprint{1203.2042})

\bibitem{Schutzhold:2008pz}
Schutzhold R, Gies H and Dunne G 2008 {\em Phys. Rev. Lett.\/} {\bf 101} 130404
  (\textit{Preprint} \eprint{0807.0754})

\bibitem{Dunne:2009gi}
Dunne G~V, Gies H and Schutzhold R 2009 {\em Phys. Rev.\/} {\bf D80} 111301
  (\textit{Preprint} \eprint{0908.0948})

\bibitem{Damgaard:1987rr}
Damgaard P~H and Huffel H 1987 {\em Phys. Rept.\/} {\bf 152} 227

\bibitem{Kim:2008yt}
Kim S~P, Lee H~K and Yoon Y 2008 {\em Phys. Rev.\/} {\bf D78} 105013
  (\textit{Preprint} \eprint{0807.2696})

\bibitem{Dunne:1998ni}
Dunne G~V and Hall T 1998 {\em Phys. Rev.\/} {\bf D58} 105022
  (\textit{Preprint} \eprint{hep-th/9807031})

\bibitem{Landsman:1986uw}
Landsman N and van Weert C 1987 {\em Phys. Rept.\/} {\bf 145} 141

\bibitem{Schmidt:1998zh}
Schmidt S, Blaschke D, Ropke G, Prozorkevich A, Smolyansky S {\em et~al.\/}
  1999 {\em Phys. Rev.\/} {\bf D59} 094005 (\textit{Preprint}
  \eprint{hep-ph/9810452})

\bibitem{Bloch:1999eu}
Bloch J~C, Mizerny V, Prozorkevich A, Roberts C~D, Schmidt S {\em et~al.\/}
  1999 {\em Phys. Rev.\/} {\bf D60} 116011 (\textit{Preprint}
  \eprint{nucl-th/9907027})

\bibitem{Tanji:2008ku}
Tanji N 2009 {\em Annals Phys.\/} {\bf 324} 1691--1736 (\textit{Preprint}
  \eprint{0810.4429})

\bibitem{Nikishov:1970br}
Nikishov A 1970 {\em Nucl. Phys.\/} {\bf B21} 346--358

\bibitem{Narozhnyi:1970uv}
Narozhnyi N and Nikishov A 1970 {\em Yad.Fiz.\/} {\bf 11} 1072

\bibitem{Kim:2008em}
Kim S~P, Lee H~K and Yoon Y 2009 {\em Phys. Rev.\/} {\bf D79} 045024
  (\textit{Preprint} \eprint{0811.0349})

\end{thebibliography}

\end{document}